\documentclass[10pt]{article}

\usepackage[utf8]{inputenc}
\usepackage[a4paper,lmargin=1.5in, rmargin=1.5in, footnotesep=1cm plus 4pt minus 4pt]{geometry}
\usepackage{setspace}
\usepackage{amsmath,amsthm,amssymb,esint}
\usepackage{mathtools, nccmath}
\allowdisplaybreaks
\usepackage{multirow}
\usepackage{sansmath}
\usepackage{pgfplots}
\pgfplotsset{compat = newest}
\usepackage{listings}
\usepackage{color}
\usepackage{placeins}
\usepackage{multicol}
\usepackage{multirow}
\usepackage{enumitem}
\usepackage{lipsum}
\usepackage{cases}
\usepackage{epstopdf}
\usepackage{float}
\usepackage{booktabs}
\usepackage{dirtytalk}
\usepackage{subcaption}
\usepackage{graphicx}

\usepackage{comment}

\usepackage[T1]{fontenc}
\usepackage{lmodern}
\usepackage{times}

\usepackage{sectsty}
\allsectionsfont{\sffamily\bfseries}
\subsubsectionfont{\sffamily\bfseries}

\usepackage[font=sf,labelfont=bf]{caption}

\usepackage[font=sf]{caption}

\newtheoremstyle{mystyle1}%                % Name
  {}%                                     % Space above
  {}%                                     % Space below
  {\itshape}%                                     % Body font
  {}%                                     % Indent amount
  {\sffamily\bfseries}%                            % Theorem head font
  {{\sffamily\mdseries .}}%                                    % Punctuation after theorem head
  { }%                                    % Space after theorem head, ' ', or \newline
  {}%                                     % Theorem head spec (can be left empty, meaning `normal')
\theoremstyle{mystyle1}

\newtheoremstyle{mystyle2}%                % Name
  {}%                                     % Space above
  {}%                                     % Space below
  {}%                                     % Body font
  {}%                                     % Indent amount
  {\itshape\sffamily}%                            % Theorem head font
  {\textsf{.}}%                                    % Punctuation after theorem head
  { }%                                    % Space after theorem head, ' ', or \newline
  {}%                                     % Theorem head spec (can be left empty, meaning `normal')
\theoremstyle{mystyle2}

\usepackage{hyperref}
\usepackage{xcolor}
\definecolor{rsrs}{RGB}{19, 59, 123}
\definecolor{myred}{rgb}{0.7, 0.0, 0.0}
\definecolor{mygreen}{rgb}{0.0, 0.3, 0.0}
\hypersetup{
    colorlinks,
    linkcolor={rsrs},
    citecolor={rsrs},
    urlcolor={rsrs}
}

\usepackage{tocloft}

\usepackage[nottoc]{tocbibind}

\usepackage{kantlipsum}

\usepackage[
    backend=biber,
    citestyle=authoryear-icomp,
    bibstyle=authoryear,
    natbib=true,
    maxbibnames=99,
    maxcitenames=2,
    giveninits=true,
    terseinits=true,
    uniquelist=false
]{biblatex}
\DefineBibliographyStrings{english}{%
    andothers = {\em et\addabbrvspace al.}
}
\DeclareNameAlias{author}{last-first}
\DeclareFieldFormat*{title}{#1}
\renewbibmacro{in:}{}

\citetrackerfalse
\DefineBibliographyStrings{english}{%
  mathesis = {MSc thesis},
}

\usepackage[title]{appendix}

\usepackage{authblk}

\usepackage[hang]{footmisc}
\usepackage[ruled, algo2e]{algorithm2e} 
\usepackage{algpseudocode}
\SetKw{KwInitialization}{Initialize}
\SetKw{KwInput}{Input:}
\SetKw{KwOutput}{Output:}

\usepackage{changepage}

\usepackage{booktabs}
\title{Exjobb}
\begin{document}

\title{ {\sffamily Probabilistic feature extraction, dose statistic prediction and dose mimicking for automated radiation therapy treatment planning} }
\author[$*$\textsf{,}$\dagger$]{Tianfang Zhang}
\author[$\dagger$]{Rasmus Bokrantz}
\author[$*$]{Jimmy Olsson}
{
\affil[$*$]{Department of Mathematics, KTH Royal Institute of Technology, Stockholm SE-100 44, Sweden}
\affil[$\dagger$]{RaySearch Laboratories, Sveavägen 44, Stockholm SE-103 65, Sweden}
}
\date{\textsf{July 6, 2021}}
\maketitle

\begin{quote}
{\centering
\section*{Abstract}
}
\textit{Purpose}: We propose a general framework for quantifying predictive uncertainties of dose-related quantities and leveraging this information in a dose mimicking problem in the context of automated radiation therapy treatment planning.

\noindent \textit{Methods}: A three-step pipeline, comprising feature extraction, dose statistic prediction and dose mimicking, is employed. In particular, the features are produced by a convolutional variational autoencoder and used as inputs in a previously developed nonparametric Bayesian statistical method, estimating the multivariate predictive distribution of a collection of predefined dose statistics. Specially developed objective functions are then used to construct a probabilistic dose mimicking problem based on the produced distributions, creating deliverable treatment plans. 

\noindent \textit{Results}: The numerical experiments are performed using a dataset of $94$ retrospective treatment plans of prostate cancer patients. We show that the features extracted by the variational autoencoder capture geometric information of substantial relevance to the dose statistic prediction problem and are related to dose statistics in a more regularized fashion than hand-crafted features. The estimated predictive distributions are reasonable and outperforms a non--input-dependent benchmark method, and the deliverable plans produced by the probabilistic dose mimicking agree better with their clinical counterparts than for a non-probabilistic formulation.

\noindent \textit{Conclusions}: We demonstrate that prediction of dose-related quantities may be extended to include uncertainty estimation and that such probabilistic information may be leveraged in a dose mimicking problem. The treatment plans produced by the proposed pipeline resemble their original counterparts well, illustrating the merits of a holistic approach to automated planning based on probabilistic modeling. 
\newline
\begin{spacing}{0.9}
{\sffamily\small \noindent \textbf{Keywords:} Knowledge-based planning, uncertainty modeling, dose--volume histogram prediction, variational autoencoder, mixture-of-experts, dose mimicking.}
\end{spacing}
\end{quote}

\tolerance=1000

\section{Introduction}

In light of its rapid advancement of late, machine learning has been extensively applied to various areas within biomedical engineering, often with considerable success \citep{park, siddique}. One such area is that of automated radiation therapy treatment planning, where data-driven approaches may help in homogenizing the otherwise time-consuming process of creating clinically satisfactory treatment plans. Instead of the conventional approach of continually adjusting weights of objective functions and possibly reiterating multiple times with physicians, plan optimization based on machine learning often comprises predicting the achievable dose-related quantities---for example, spatial dose or dose--volume histograms (DVHs)---and solving the inverse problem of reconstructing them with respect to machine parameters. While much of previous literature focuses on pure prediction, the quantification of its associated uncertainty, which also contains valuable information for the subsequent optimization step, is often omitted. To address this, we present a general framework for, given any collection of dose-related quantities, estimating their multivariate predictive probability distribution and setting up a corresponding plan optimization problem taking this probabilistic information into account. 

In general, automated treatment planning using machine learning, also known as knowledge-based planning, concerns the automatic plan generation using knowledge extracted from historically delivered clinical treatment plans \citep{ge, ng, wang, hussein}. One usually goes about this task by first assuming a parameterized optimization problem and then training a machine learning model to predict the unknown parameters, which is done in such a way that the solution to the resulting optimization problem will correspond to a clinically satisfactory plan. An example is the prediction of weights in a weighted-sum formulation \citep{boutilier}, which may even be sequentially adjusted by an agent trained using reinforcement learning methods \citep{shen}. More commonly, one predicts either a spatial dose distribution, a set of DVHs in some regions of interest, a collection of dose statistics or a combination thereof, and then minimizes the deviation between the quantities evaluated on the actual dose and their corresponding predicted values. This latter optimization is often referred to as dose mimicking. The prediction--mimicking division is prevalent in large parts of the current literature on knowledge-based planning, allowing for the parts to be separately investigated but also leading to a certain lack of causality between prediction accuracy and resulting plan quality.

The first part of predicting dose-related quantities has been abundantly addressed in previous literature. Regarding spatial dose prediction, two- or three-dimensional convolutional neural networks in various architectures, such as U-nets, have been widely used \citep{nguyen_unet, campbell, kearney, shiraishi, ma_isodose} and extended to generative models such as generative adversarial networks \citep{babier_gan, murakami}, although other methods such as random forests have also been studied \citep{mcintosh}. For prediction of DVHs or other dose statistics, while overlap volume histograms evaluated on the input image have been traditionally used for this purpose \citep{appenzoller, wall, jiao, skarpmanmunter, ma_features, zhu, wu, yuan}, more recent literature also includes the use of neural network--based methods to simultaneously predict spatial dose and DVHs directly from input images \citep{liu, nguyen_pareto}. Common for all of the aforementioned approaches, however, is that predictions are made deterministically with no associated predictive probability distribution. Various sources for inter-planner variations are investigated in \citet{nelms}, and as pointed out by \citet{babier_importance}, when designing and evaluating methods for automated planning, it is essential to maintain a holistic picture of both parts of a prediction--mimicking pipeline. Considering this, omitting the probabilistic information contained in predictive distributions when using purely predictive methods may be the cause of a substantial disconnect between the two parts. For example, the predictive uncertainty associated with the mean dose in some organ at risk may be orders of magnitude higher than that of the near-minimum dose in the planning target volume (PTV), which is crucial information for the dose mimicking problem; the same holds for individual voxel doses in spatial dose prediction. Without this information, finding the right weights in an optimization problem minimizing deviations between DVH statistics or voxelwise doses may be just as hard as tuning objective weights in conventional inverse planning. 

While certain aspects of a general probabilistic approach have been previously addressed---for instance, in \citet{nilsson}, where a U-net--based mixture density network is trained to output voxelwise univariate Gaussian mixtures, or in \citet{nguyen_bagging}, where ensemble and Monte Carlo dropout techniques are used to provide uncertainties of spatial dose and dose statistics by sampling predictions---the estimation of concrete predictive distributions for DVHs or other dose statistics has yet to be given much attention in literature. In particular, given a collection of dose statistics, which may represent a discretized DVH, one would like to estimate their multivariate conditional probability distribution given the current patient geometry and the training dataset. For example, Eclipse RapidPlan (Varian Medical Systems, Palo Alto, CA, USA) provides DVH prediction uncertainties by forming confidence bands based on estimated standard deviations around the predictive mean for each point on the DVH \citep{fogliata}, although the actual underlying probability distribution is not available. As the raw inputs are typically contoured CT images (or, alternatively, only the contours), prediction of dose statistics or DVHs is a high-dimensional problem made especially hard by the scarcity of data often prevalent in similar settings. One way to remedy this problem is to employ dimensionality reduction using unsupervised learning methods to preprocess the input images before passing them to a machine learning model. Such dimensionality reduction methods have been previously explored in e.g. \citet{hanna}, where a variational autoencoder \citep{kingma} is used for the purpose, but also in other related areas, examples including deformable registration \citep{fu} and outcome prediction \citep{cui}.

In this paper, we propose a general probabilistic framework for quantifying predictive uncertainties of any collection of dose-related quantities and leveraging the information for creating a deliverable plan. This is done in a three-step pipeline comprising the extraction of features, the prediction of dose statistics and the solving of a dose mimicking problem, each of which enjoying the advantage of being substitutable by any equivalent method. In particular, we use a convolutional variational autoencoder for the dimensionality reduction of contoured CT images, producing features to be used as inputs in a recently developed nonparametric Bayesian method \citep{zhang_sbmoe} exploiting feature similarities and outputting predictive distributions as Gaussian mixtures. The estimation of such joint predictive distributions incorporating dependencies between dose statistics is novel in literature, as is the use of variational autoencoders for producing covariates to predict dose-related quantities. Furthermore, using specially developed objective functions designed to leverage the information contained in the estimated distributions, with gradients of dose statistics readily available using the method outlined in \citet{zhang_direct}, a probabilistic dose mimicking formulation is solved to produce complete treatment plans. The computational study shows that the proposed pipeline yields extracted features containing geometric information of substantial relevance for dose statistic prediction, reasonable estimations of multivariate predictive distributions over dose statistics and complete treatment plans agreeing well with their respective ground truths. More specifically, the variational autoencoder is shown to produce a less entangled feature space than that of hand-crafted features, the estimated predictive distributions outperform a non--input-dependent benchmark method, and the probabilistic dose mimicking formulation is shown to be more suited than a non-probabilistic analogue for handling planning tradeoffs. In conclusion, the experiments serve to demonstrate the feasibility of the framework for automated treatment planning.

\section{Materials and methods}

Let $\{(x^n, d^n)\}_{n} \subset \mathcal{X} \times \mathcal{D}$ be a dataset of clinical, historically delivered treatment plans consisting of pairs of contoured CT images $x^n$ and dose distributions $d^n$, which are represented as vectors $x^n = (x_i^n)_i$ and $d^n = (d_i^n)_i$, $i$ being the index over voxels. Here, $\mathcal{X}$ and $\mathcal{D}$ denote spaces of contoured images and dose distributions, respectively. For our purposes, we will only use the binary encodings of the regions of interest (ROIs) and not the radiodensities in the CT image. Let also $\{\psi_j\}_{j}$ be a collection of dose statistic functions $\psi_j : \mathcal{D} \to \mathbb{R}$---e.g., dose-at-volume $\operatorname{D}_v$ in different ROIs and at different volume levels $v$---and let $y^n$ be defined as the vector $y^n = (\psi_j(d^n))_{j}$ of evaluated dose statistic values on $d^n$ for each $n$. In the following, we will use the terms dose statistic and dose-related quantity interchangeably. 

\begin{figure}[H]
\centering
\includegraphics[width=0.9\textwidth]{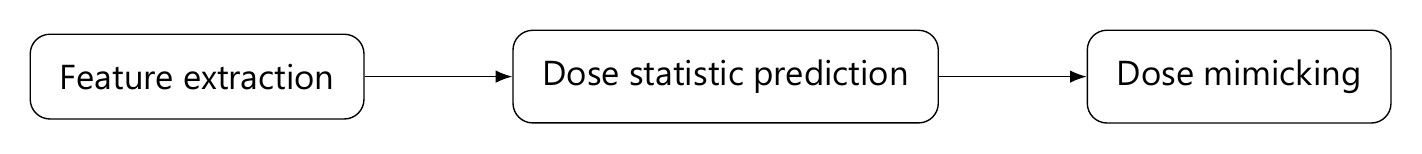}
\caption{The proposed automatic treatment planning pipeline.}
\label{flowchart}
\end{figure}

Given a new patient with input image $x^*$, the main task is to predict the corresponding dose statistic values $y^*$ and solve an accordingly constructed dose mimicking optimization problem. Our pipeline, shown in Figure \ref{flowchart}, is divided into the following three parts:
\begin{enumerate}
    \item training a variational autoencoder to extract features $\phi(x^*) \in \mathcal{Z}$ for some relatively low-dimensional vector space $\mathcal{Z}$, where $\phi$ is the encoder part,
    \item using $\{(\phi(x^n), y^n)\}_{n}$ as training set to train a similarity-based Bayesian mixture-of-experts model, developed in a previous paper \citep{zhang_sbmoe}, which outputs an estimate of the multivariate predictive density $p(y^* \mid x^*, \{(x^n, y^n)\}_n)$, and
    \item solving a dose mimicking problem using specially developed objective functions incorporating probabilistic information in the predictive density $p(y^* \mid x^*, \{(x^n, y^n)\}_n)$, thus creating a deliverable plan. 
\end{enumerate}

\subsection{Variational autoencoder}
\label{vae}

Using neural networks to extract features from high-dimensional images is a form of representation learning \citep{bengio}. A common method for this is using autoencoders \citep{goodfellow}, which is an unsupervised learning method using only the training inputs $x^n$. An autoencoder consists of an encoder $\phi : \mathcal{X} \to \mathcal{Z}$ and a decoder $\varphi : \mathcal{Z} \to \mathcal{X}$, and is trained in such a way that $\varphi(\phi(x^*))$ will resemble each new input $x^*$ as well as possible. If this is successful, one can deem the low-dimensional feature vector $\phi(x^*)$ to contain sufficient input to reconstruct $x^*$ reasonably well, thereby making $\phi$ qualify as a feature extractor.

A drawback with such plain autoencoders is the possible lack of regularity---that is, there is generally no guarantee that the decoded images $\varphi(z)$, $\varphi(z')$ will be close to each other whenever $z$, $z'$ are close. This may be problematic when using the produced features in another statistical model, particularly if the model is a nonparametric model relying on interpolations. While regularization methods for autoencoders have been extensively studied \citep{bengio}, another way of approaching this problem is to instead use a variational autoencoder, replacing the encoder and decoder by probabilistic counterparts. Modeling the latent representation $z \in \mathcal{Z}$ corresponding to each input $x \in \mathcal{X}$ as a random variable, in a variational autoencoder, one uses neural networks to approximate the conditional densities $p(z \mid x)$ and $p(x \mid z)$. The latter density, referred to as the likelihood, is often well-defined whereas the former, known as the posterior, is subsequently derived from Bayes' rule if one also defines a prior $p(z)$. However, the posterior $p(z \mid x)$ in this case will be intractable to compute---instead, it is approximated with some closed-form density $q(z)$, often called the variational posterior, in such a way that the Kullback--Leibler divergence $d_{\mathrm{KL}}(q \; \Vert \; p(\cdot \mid x))$ between $q(z)$ and $p(z \mid x)$ is minimized. Equivalently, one may find $q(z)$ by maximizing the evidence lower bound $\operatorname{ELBO}(q) = \operatorname{\mathbb{E}}^{q(z)} \log(p(x, z) / q(z))$. For a detailed review of variational inference in general, see \citet{blei}.

In our case, given a fixed set of ROIs $\{R_k\}_{k=1}^K$ as index sets over voxels (each voxel may belong to several ROIs), a binary encoding of each input image $x = (x_i)_i \in \mathcal{X}$ is used, where $x_i$ is the vector $x_i = (1_{i \; \in \; R_k})_k \in \mathbb{R}^{K}$---that is, entry $k$ of $x_i$ equals $1$ if voxel $i$ belongs to ROI $R_k$ and $0$ otherwise. For the model, we follow \citet{kingma} and define for each $n$ the likelihood and prior
\[
p(x^n \mid z^n) = \prod_i \prod_k \operatorname{Be}\!\left( x_{ik}^n \mid \varphi(z^n)_{ik} \right), \quad p(z^n) = \prod_j \operatorname{N}(z^n_j \mid 0, 1),
\]
where $z^n = (z^n_j)_j$, with the variational posterior being
\[
q(z^n) = \prod_j \operatorname{N}(z^n_j \mid \phi(x^n)_{j1}, \phi(x^n)_{j2}^2)
.\]
Here, $\operatorname{Be}$ and $\operatorname{N}$ denote the Bernoulli and the normal distribution, and $\phi$ and $\varphi$ are two neural networks with outputs of shape $\operatorname{dim} \mathcal{Z} \times 2$ and $\operatorname{dim} \mathcal{X} \times K$, respectively. The negative ELBO, which is the loss function to be minimized with respect to the network weights in $\phi$ and $\varphi$, is then obtained as
\begin{align*}
-\operatorname{ELBO}(q) = \sum_n \Bigg(& \frac{1}{2} \sum_j (1 + \log \phi(x^n)_{j2}^2 - \phi(x^n)_{j1}^2 - \phi(x^n)_{j2}^2) \\
\quad\quad &- \operatorname{\mathbb{E}}^{q(z^n)} \sum_i \sum_k \Big( x_{ik} \log \varphi(z^n)_{ik} + (1 - x_{ik}) \log(1 - \varphi(z^n)_{ik}) \Big) \Bigg)
\end{align*}
---here, the sum over $j$ regularizes the variational posterior by penalizing discrepancy from the prior, while the expected sum over $i$ and $k$ penalizes reconstruction error. In particular, the expectations over $q(z^n)$ and their gradients may be Monte Carlo--approximated using the reparameterization trick \citep{kingma}, which relies on rewriting $z^n$ as a location--scale transform $z^n = \phi(x^n)_{j1} + \epsilon \phi(x^n)_{j2}$ with $\epsilon \sim \operatorname{N}(0, 1)$ and then sampling $\epsilon$ at each objective evaluation.

\subsection{Dose statistic prediction}

Having trained the variational autoencoder, we may treat the encoder as fixed and use the set $\{(\phi(x^n), y^n)\}_n$, comprising pairs of feature vectors and evaluated dose statistics, as training dataset for a statistical model in which the multivariate predictive distribution $p(y^* \mid \phi(x^*), \{(\phi(x^n), y^n)\}_n)$ is obtained given a new input $x^*$. However, despite the embedding of the inputs as feature vectors, the dimensionalities involved are still high relative to typical dataset sizes. Due to the rather complex relationships between feature vectors and dose statistics, this will make the training of a parametric model of suitable size---e.g., a neural network outputting for each input a multivariate normal distribution predicting the corresponding output---hard to train to acceptable generalization accuracy without overfitting. Moreover, as distributions of dose statistics may be skewed or even multimodal \citep{nilsson}, such a model will often be inadequate even if it were possible to train using the available data.

Motivated by this, we settle on a so called similarity-based mixture-of-experts model \citep{zhang_sbmoe} for the purpose, which has predictive distributions in the form of multivariate Gaussian mixtures. This is a recently developed Bayesian nonparametric model in which predictions are made based on similarities between inputs rather than explicitly modeling input--output relationships. In particular, for each new input feature vector $\phi(x^*)$, mixture weights are calculated by first evaluating the similarity $\tau_n$ to each equivalent $\phi(x^n)$ in the training set, and then the probability $\sigma_{nc}$ of each $y^n$ belonging to each expert class $c$---the $\tau_n$ are referred to as first transitions and the $\sigma_{nc}$ as second transitions. The experts, in turn, are multivariate normal distributions $\{\operatorname{N}(\mu_c, \Sigma_c)\}_{c=1}^C$. Specifically, the distance metric $d_{\mathcal{Z}}$ used to determine similarities in the feature space $\mathcal{Z}$ is the Mahalanobis distance $d_{\mathcal{Z}}(z, z') = (z - z')^{\operatorname{T}} \Lambda (z - z')$, where $\Lambda$ is a precision matrix. This leads to a predictive likelihood on the form
\[
p(y^* \mid \phi(x^*), \{(\phi(x^n), y^n)\}_n, \theta) = \sum_c \sum_n \tau_n \sigma_{nc} \operatorname{N}(y^* \mid \mu_c, \Sigma_c)
,\]
where $\theta = (\Lambda, \{(\mu_c, \Sigma_c)\}_c)$ are the parameters and where the first and second transitions $\tau_n$, $\sigma_{nc}$ are evaluated as
\[
\tau_n = \frac{\operatorname{N}(\phi(x^*) \mid \phi(x^n), \Lambda^{-1})}{\sum_{n'} \operatorname{N}(\phi(x^*) \mid \phi(x^{n'}), \Lambda^{-1})}, \quad \sigma_{nc} = \frac{\operatorname{N}(y^n \mid \mu_c, \Sigma_c)}{\sum_{c'} \operatorname{N}(y^n \mid \mu_{c'}, \Sigma_{c'})}
.\]
The parameters $\theta$ are treated in a Bayesian fashion, fitted using updates according to a mean-field variational Bayes algorithm, and the resulting predictive distribution is obtained from Monte Carlo samples from the variational posterior. For further details, see \citet{zhang_sbmoe}.

\subsection{Dose mimicking}
\label{dosemimickingsec}

\begin{figure}[h]
\centering
\includegraphics[width=0.9\textwidth]{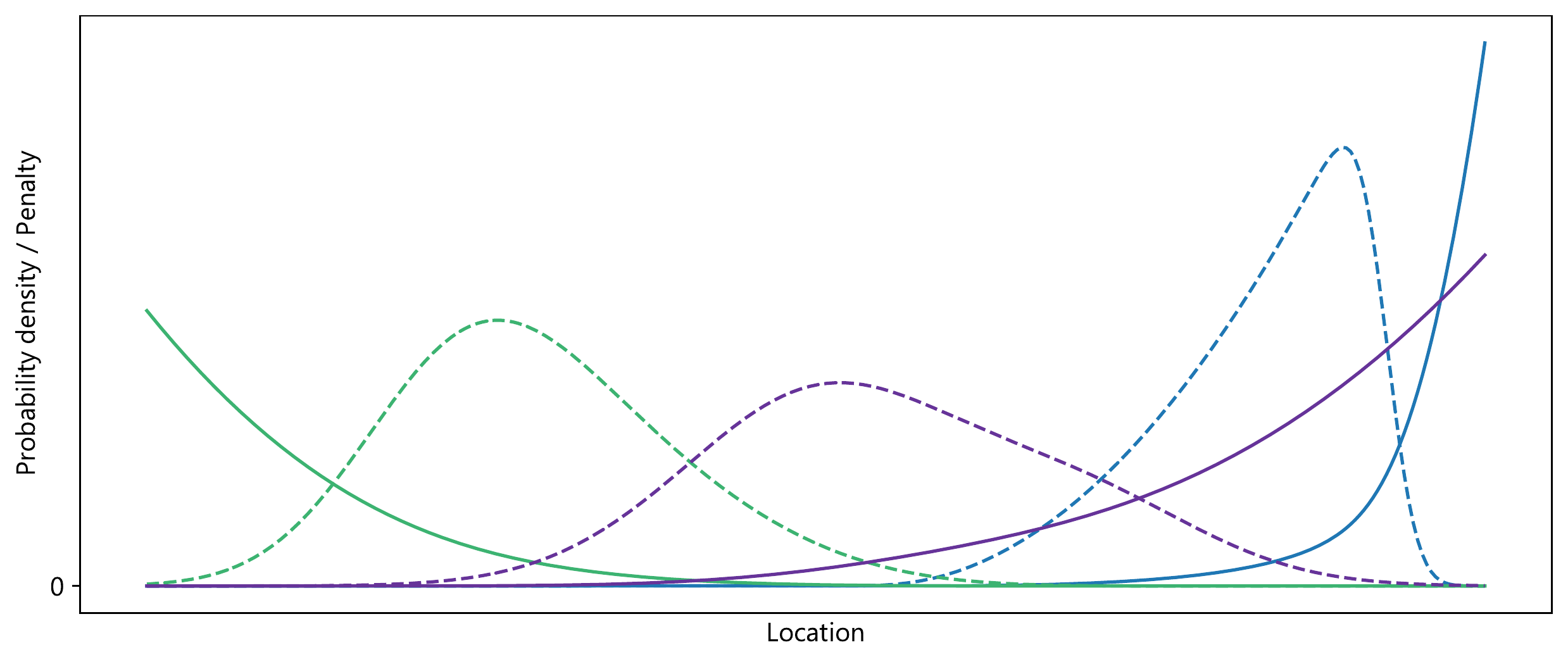}
\caption{Illustration of how the shapes of the predictive densities (dashed) affect their respective penalty contributions (solid). The blue and purple lines correspond to dose statistics to be minimized and the green lines to those to be maximized.}
\label{tce}
\end{figure}

With a feature extraction method and dose statistic prediction model in place, consider a new patient with input image $x^*$ for which the predictive distribution $p(y^* \mid x^*, \{(x^n, y^n)\}_n)$ over a vector $y^* = (\psi_j(d^*))_j$ of dose statistic values has been estimated. In particular, let $F_j$ be the cumulative distribution function of the univariate marginal density $p(y^*_j \mid x^*, \{(x^n, y^n)\}_n)$ for each $j$. To utilize the probabilistic information captured in these distributions, we will base the objective function contribution from dose statistic $j$ for a given dose $d$ on the value of $F_j(\psi_j(d))$---loosely speaking, the contribution is based on the fraction of training patients the current patient is better than in terms of dose statistic $j$. While one could, in principle, use the joint cumulative distribution function over all dose statistics, the choice of using the marginal distributions separately is motivated by the advantage of increased control over individual dose statistics. We assume that each dose statistic is either ideally maximized or minimized, and assign binary labels $t_j = 1$ and $t_j = 0$ for the former and the latter case, respectively, to reflect this. Since the cumulative distribution function ranges from $0$ to $1$, we set up the loss on a cross-entropy form to obtain the contribution corresponding to $\psi_j$ as
\begin{equation}
\label{ce_objective}
\operatorname{CE}_j(d) = -t_j \log F_j(\psi_j(d)) - (1 - t_j) \log(1 - F_j(\psi_j(d))).
\end{equation}
Note that this means that the higher the certainty that the dose statistic is around some range of values, the more will the objective penalize deviations from those values. At the same time, since $F_j$ is continuous and strictly increasing in the case of a mixture-Gaussian predictive density, the optimizer will always have incentive to improve even when beyond the range of typical values. Figure \ref{tce} illustrates the behavior of the penalty for different probability distributions.

As for the resulting optimization problem, let $\eta$ denote the optimization variables with feasible set $\mathcal{E}$, where the total dose $d$ is determined by some dose deposition mapping $d = d(\eta)$. For example, in the case of direct machine parameter optimization for volumetric modulated arc therapy (VMAT), the relation between the machine parameters $\eta$ and the dose $d$ may be modeled according to \citet{unkelbach}. The optimization problem may then be written as
\begin{equation}
\label{optformulation}
\underrel{\text{minimize}}{\eta \in \mathcal{E}} \quad \sum_j w_j \operatorname{CE}_j(d(\eta)),
\end{equation}
where $w_j$ is an importance weight for each dose statistic $\psi_j$. It is important to note that in contrast to conventional inverse planning or dose mimicking, the optimization problem above is relatively insensitive to the weights $w_j$ since a substantial part of the information regarding the relative importances of the dose statistics is already stored in the cumulative distribution functions $F_j$. 

\subsection{Computational study}

To showcase the advantages and disadvantages of the proposed approach, we will use the following numerical experiments to demonstrate the full pipeline comprising feature extraction, dose statistic prediction and dose mimicking. The data used for this purpose consists of $94$ retrospective treatment plans of prostate cancer patients having undergone a prostatectomy prior to radiation therapy, originating from the Iridium Cancer Network in Antwerp, Belgium. The ROIs were contoured according to the RTOG \citep{gay} and ACROP \citep{salembier} guidelines, and the patients were treated with a prescribed dose of $7000 \; \mathrm{cGy}$ in the prostate bed and $5600 \; \mathrm{cGy}$ in the seminal vesicles and pelvic nodes, divided into $35$ fractions. All patients were treated using dual $360$-degree VMAT arcs, with final doses calculated using a collapsed cone algorithm. The DVHs of the training patients are shown in the backgrounds of Figure \ref{dvhband_1}. For the feature extraction and dose statistic prediction parts, the dataset was split into a training and a test set of $84$ and $10$ patients, respectively.

\begin{figure}[h]
\hspace*{-1.8cm}
\includegraphics[width=1.3\textwidth]{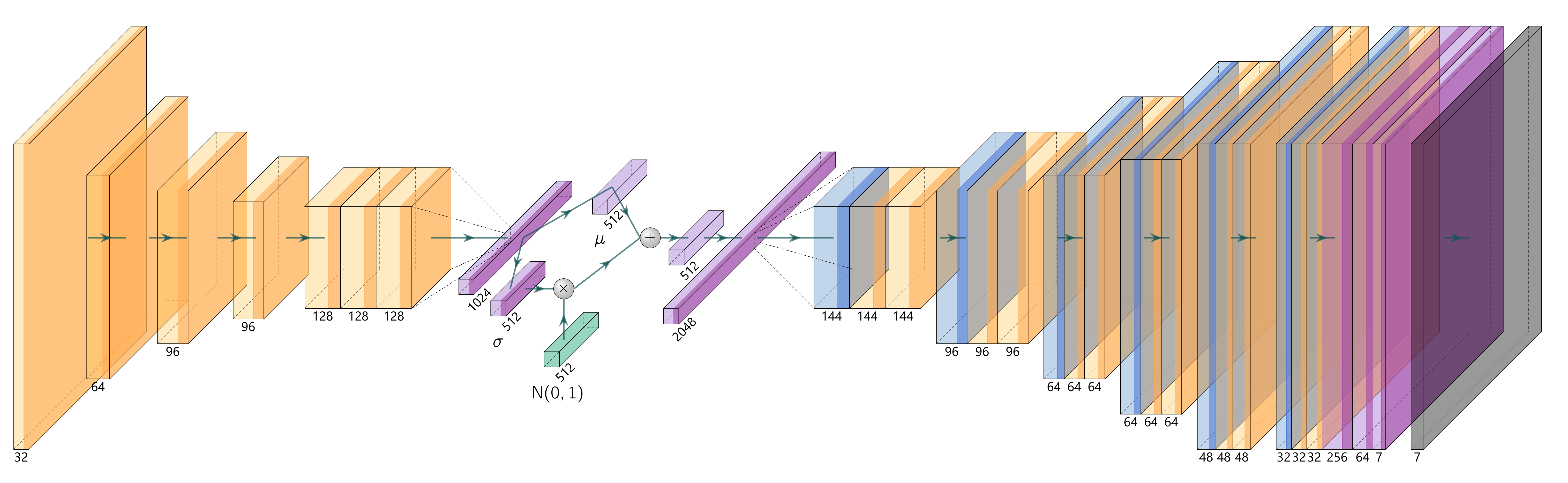}
\caption{Illustration of the architecture of the variational autoencoder described in Section \ref{vae}. The encoder consists of convolutional (yellow) and dense (purple) layers, and the decoder upsamples using transpose-convolutional (blue) layers. After each layer, a ReLU activation is applied if indicated with a darker color, followed by layer normalization and dropout. The green vector represents the stochastic source $\epsilon$ in the reparameterization trick described in Section \ref{vae}, and the last gray layer represents a softmax activation.}
\label{vae_architecture}
\end{figure}

The variational autoencoder described in Section \ref{vae} was implemented in TensorFlow 2.3, using the architecture depicted in Figure \ref{vae_architecture}. Specifically, convolutional and transpose-convolutional layers were mostly used for the encoder and decoder parts, although complemented with dense layers, and layer normalization and dropout were applied after each rectified linear unit (ReLU) activation. Using the prostate PTV, seminal vesicles PTV, rectum, bladder, left femur, right femur and small bowel as the set of ROIs whose contours are inputted to the neural network---a total of seven ROIs---the input images were all preprocessed to binary arrays of size $64 \times 48 \times 83 \times 7$ with a $5 \; \mathrm{mm}$ voxel resolution, and the dimension of the latent space $\mathcal{Z}$ was set to $512$. In total, the network comprised around $6$ million parameters, which were fitted using a standard Adam optimizer \citep{kingma_adam} and a training and validation set of $75$ and $9$ patients, respectively. In particular, random data augmentations of small shifts and rotations were employed during training. The features produced by the variational autoencoder were qualitatively compared to hand-crafted features. In our case, we used distance transforms from the rectum and bladder regions to themselves and the two PTVs represented by normalized histograms with bin edges $(-3, -2,  -1, 0, 1, 2, 3, 5, 7.5, 10) \; \mathrm{cm}$, amounting to $88$ dimensions in total.

The similarity-based mixture-of-experts model was subsequently trained using the implementation described in \citet{zhang_sbmoe}, in particular also using Tensorflow 2.3. For organs at risk, the set $\{\psi_j\}_j$ of dose statistics were dose-at-volume levels at volumes $10 \; \%, 20 \; \%, \dots, 90 \; \%$, all with $t_j = 0$; for targets, both dose-at-volume and mean-tail-dose \citep{romeijn} at volumes $2 \; \%, 5 \; \%, 10 \; \%, 90 \; \%, 95 \; \%, 98 \; \%$ were used, with $t_j = 1$ for volumes less than $50 \; \%$ and $t_j = 0$ otherwise. Here, mean-tail-dose was included to achieve better control of the tails of the target DVHs \citep{zhang_direct}. While one could in principle train one model to fit all dose statistics at once, in order to achieve better predictive accuracy given the small amount of available data, we trained one mixture-of-experts model per ROI at the cost of sacrificing inter-ROI dependencies. Each model comprised $C = 32$ experts with a posterior sample size of $10$ for each mean--covariance pair $(\mu_c, \Sigma_c)$, leading to a mixture model of $320$ classes.

The dose mimicking problem (\ref{optformulation}) was subsequently set up using the fitted dose statistic prediction model and the same beam configuration as used in the training dataset, with all optimizations and dose calculations performed in RayStation 10B (RaySearch Laboratories). In particular, the direct machine parameter optimization formulation for VMAT described in Section 2c of \citet{unkelbach} was used, leading to fully deliverable plans. To obtain increased control of the resulting plans, the clinical goals in Table \ref{tab:clinical_goals} were added to the set of dose statistics, with predictive distributions substituted by synthetically constructed normal distributions centered around the respective goal values. The importance weights $w_j$ were set to $0.1$ for the left and right femurs, $1$ for the rectum, bladder and small bowel regions, and $10$ for the PTVs. The objective functions, with gradients available due to the recent developments in \citet{zhang_direct}, were implemented in the RayStation source code. Each mimicking optimization comprised $3$ runs of $100$, $100$ and $40$ iterations with accurate dose computations in between---as per standard practice, approximate doses during optimization were calculated by a singular value decomposition algorithm and accurate doses by a collapsed cone algorithm. To illustrate the contribution of the uncertainty information captured in the cross-entropy objectives $\operatorname{CE}_j(d)$ from (\ref{ce_objective}), we also compared to a similar but non-probabilistic dose mimicking formulation. Specifically, each $\operatorname{CE}_j(d)$ in the proposed dose mimicking problem (\ref{optformulation}) was replaced by a quadratic penalty
\[
t_j (\psi_j(d) - \hat{\psi}_j)_{-}^2 + (1 - t_j) (\psi_j(d) - \hat{\psi}_j)_{+}^2
,\]
with $\hat{\psi_j}$ set to the mean of the distribution associated with $F_j$ and $(x)_-$, $(x)_+$ denoting the negative and positive parts of $x$, respectively. Otherwise, the quadratic penalty formulation used the same optimization settings and weights as the original dose mimicking procedure. 

\begin{table}[h]
\caption{\label{tab:clinical_goals}Additional dose-at-volume ($\operatorname{D}_v$) and lower/upper mean-tail-dose ($\operatorname{MTD}_v^{\pm}$) clinical goals used in the dose mimicking problem.}
\centering
\begin{tabular}{ll}
\toprule
ROI & Goal \\
\midrule
PTV, prostate & $\operatorname{MTD}^-_{98 \, \%} \geq 6700 \; \mathrm{cGy}$ \\
PTV, prostate & $\operatorname{MTD}^+_{0.5 \, \%} \leq 7400 \; \mathrm{cGy}$ \\
PTV, seminal vesicles & $\operatorname{MTD}^-_{98 \, \%} \geq 5300 \; \mathrm{cGy}$ \\
PTV, seminal vesicles & $\operatorname{MTD}^+_{5 \, \%} \leq 5870 \; \mathrm{cGy}$ \\
Rectum & $\operatorname{D}_{50 \, \%} \leq 4500 \; \mathrm{cGy}$ \\
Bladder & $\operatorname{D}_{25 \, \%} \leq 6500 \; \mathrm{cGy}$ \\
Near-target & $\operatorname{MTD}^+_{5 \, \%} \leq 5500 \; \mathrm{cGy}$ \\
\bottomrule
\end{tabular}
\label{tableresults}
\end{table}

\begin{comment}
\begin{table}[h]
    \centering
    \caption{\label{tab:clinical_goals}Additional dose-at-volume ($\operatorname{D}_v$) and lower/upper mean-tail-dose ($\operatorname{MTD}_v^{\pm}$) clinical goals used in the dose mimicking problem.}
    \begin{tabular}{l|l}
        ROI & Goal \\\hline
        PTV, prostate & $\operatorname{MTD}^-_{98 \, \%} \geq 6700 \; \mathrm{cGy}$ \\
        PTV, prostate & $\operatorname{MTD}^+_{0.5 \, \%} \leq 7400 \; \mathrm{cGy}$ \\
        PTV, seminal vesicles & $\operatorname{MTD}^-_{98 \, \%} \geq 5300 \; \mathrm{cGy}$ \\
        PTV, seminal vesicles & $\operatorname{MTD}^+_{5 \, \%} \leq 5870 \; \mathrm{cGy}$ \\
        Rectum & $\operatorname{D}_{50 \, \%} \leq 4500 \; \mathrm{cGy}$ \\
        Bladder & $\operatorname{D}_{25 \, \%} \leq 6500 \; \mathrm{cGy}$ \\
        Near-target & $\operatorname{MTD}^+_{5 \, \%} \leq 5500 \; \mathrm{cGy}$ \\
    \end{tabular}
\end{table}
\end{comment}

\section{Results}

\begin{figure}[h]
\centering
\begin{subfigure}[t]{\textwidth}
\centering
\includegraphics[width=\textwidth]{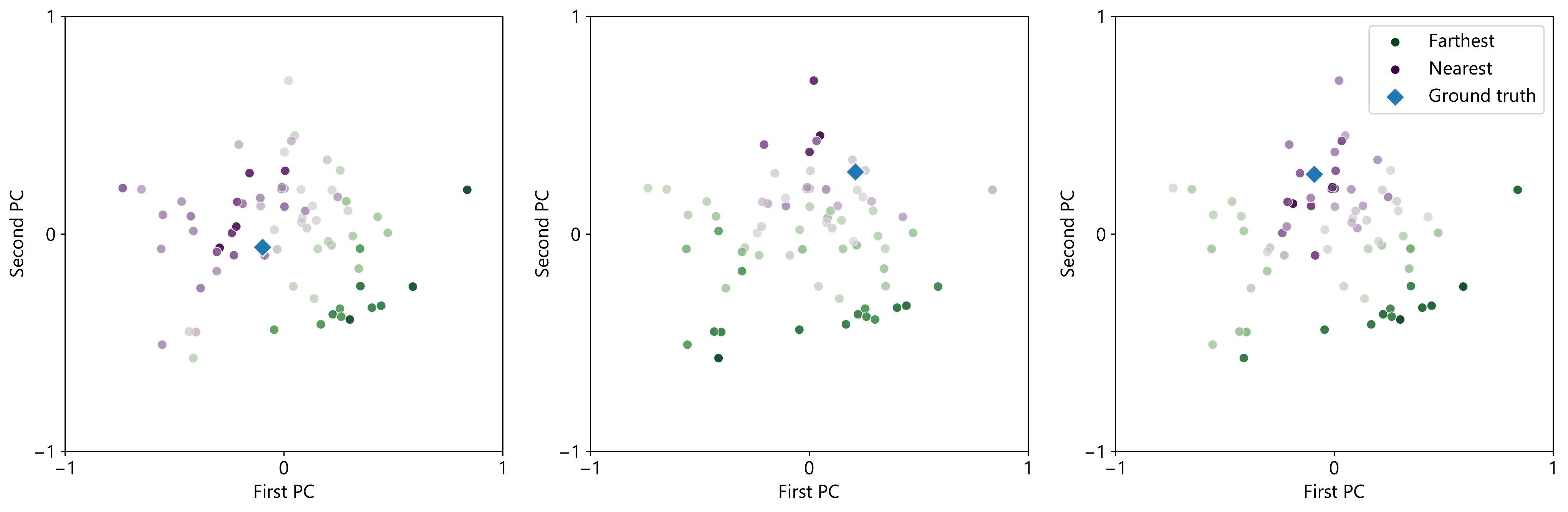}
\caption{Variational autoencoder features.}
\end{subfigure}%
\vspace{0.5cm}
\begin{subfigure}[b]{\textwidth}
\centering
\includegraphics[width=\textwidth]{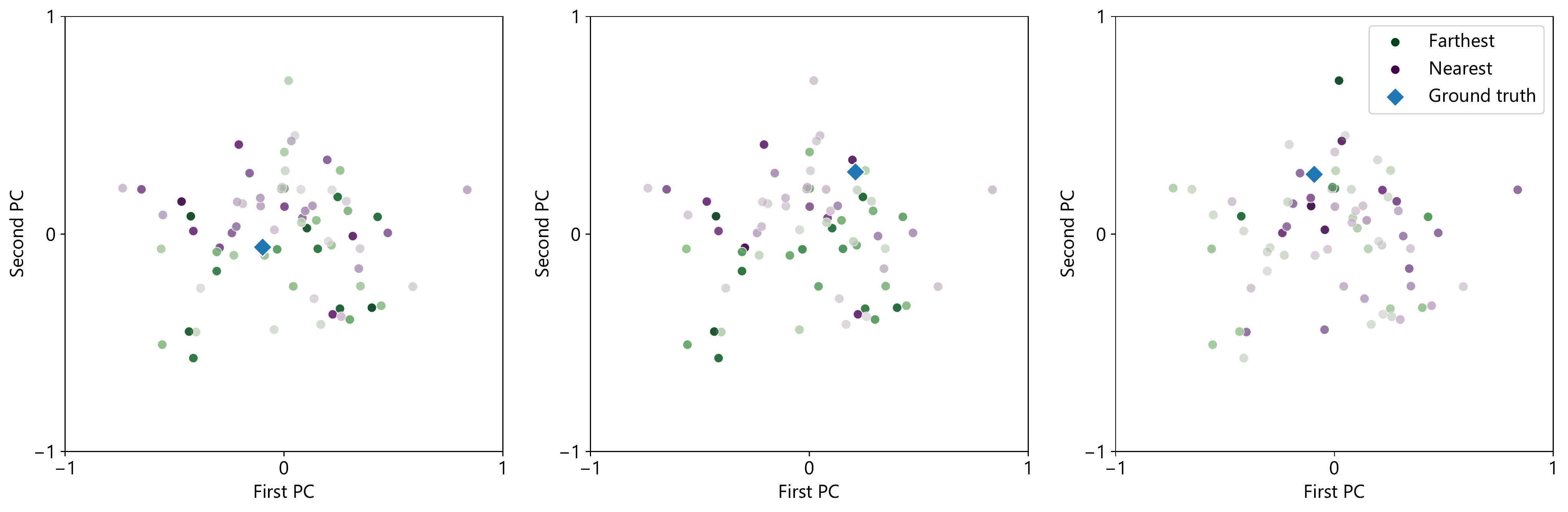}
\caption{Hand-crafted features.}
\end{subfigure}
\caption{Scatterplot of the first and second principal components (PCs) of the dose statistics in the training data (round) and one test patient (diamond), plotted for three patients in the test dataset. A darker purple (green) color corresponds to being closer (farther) in the feature space with respect to a relevance-adapted metric constructed from variational autoencoder (a) and hand-crafted (b) features.}
\label{vae_features_linear}
\end{figure}

As the main purpose of the variational autoencoder is to provide input features for the dose statistic prediction models, we first ensure that the produced features contain sufficient predictive power. For this, we temporarily use a simplistic linear regression model fitted using the data $\{(\phi(x^n), y^n_{\mathrm{pc}})\}_n$, where $y^n_{\mathrm{pc}} = (y^n_{\mathrm{pc}, 1}, y^n_{\mathrm{pc}, 2})$ are the two first principal components of the dose statistic values over all ROIs. Then, for each patient $x^*$ in the test dataset, we compute the standard Euclidean distance between the predicted $y_{\mathrm{pc}}^*$ and each predicted $y^n_{\mathrm{pc}}$, which serves as a relevance-adapted distance metric in the feature space $\mathcal{Z}$. This distance metric is also compared to one constructed analogously but using the hand-crafted distance transform features. Figure \ref{vae_features_linear} depicts this distance in relation to the actual distance between the principal component vectors in scatterplots for three test patients, comparing between the variational autoencoder and the hand-crafted features. Despite the simplicity of the linear model, for the variational autoencoder features, the relevance-adapted distances agree well with the principal component distances, with points lying close to the ground truth also being close in feature space. As a start, this shows that the features produced by the variational autoencoder contain substantial information of relevance for predicting dose statistics. Moreover, the scatterplots for the hand-crafted features show a more irregular relation between the relevance-adapted distances and the true principal component distances. While this does not per se entail that these features contain less predictive power than those produced by the variational autoencoder, it is indicative of a more entangled feature space, which may aggravate the training of the subsequent dose statistic prediction model.

\begin{figure}[h]
\centering
\includegraphics[width=\textwidth]{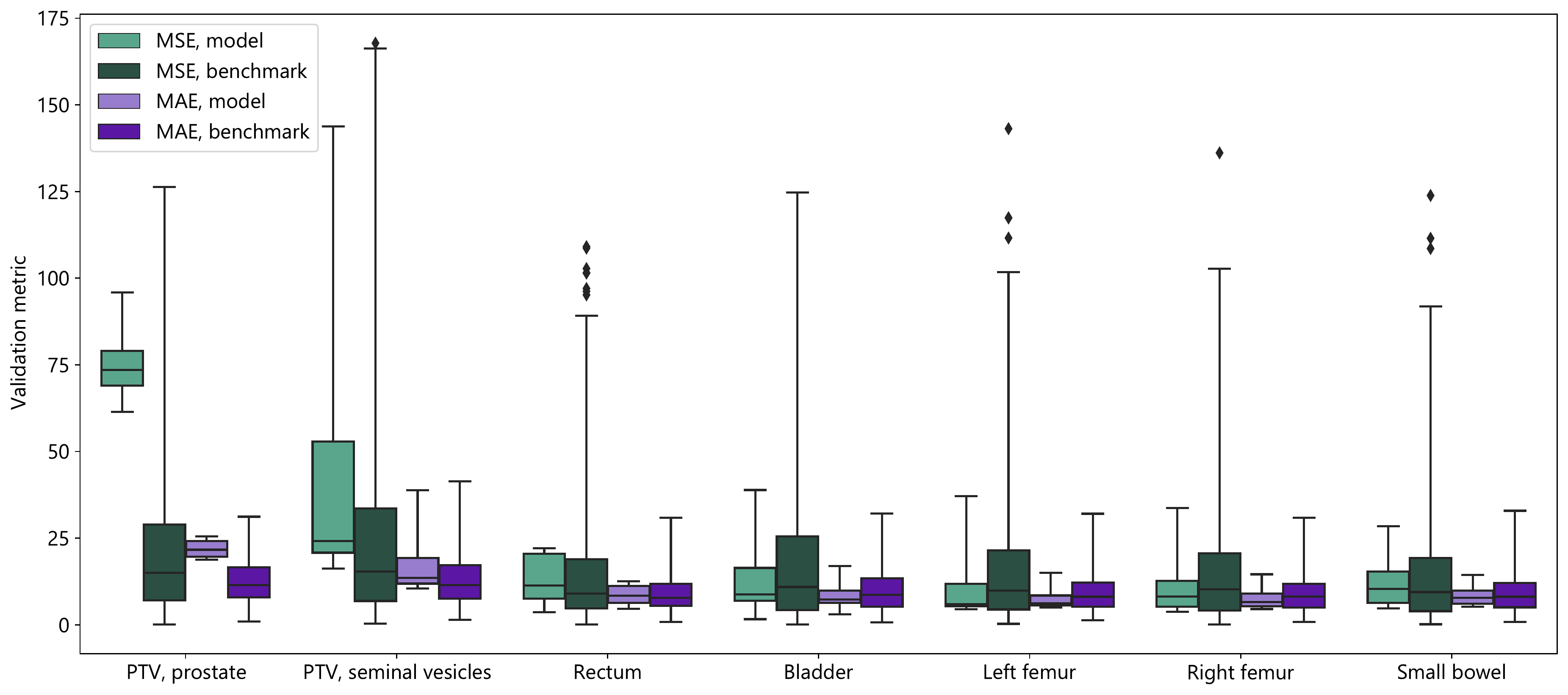}
\caption{Boxplots of the MSE and MAE for the different models in comparison to their benchmark models. The diamond-shaped markers show outliers.}
\label{metric_boxplot}
\end{figure}

\begin{figure}[h]
\centering
\begin{subfigure}[t]{\textwidth}
\centering
\includegraphics[width=\textwidth]{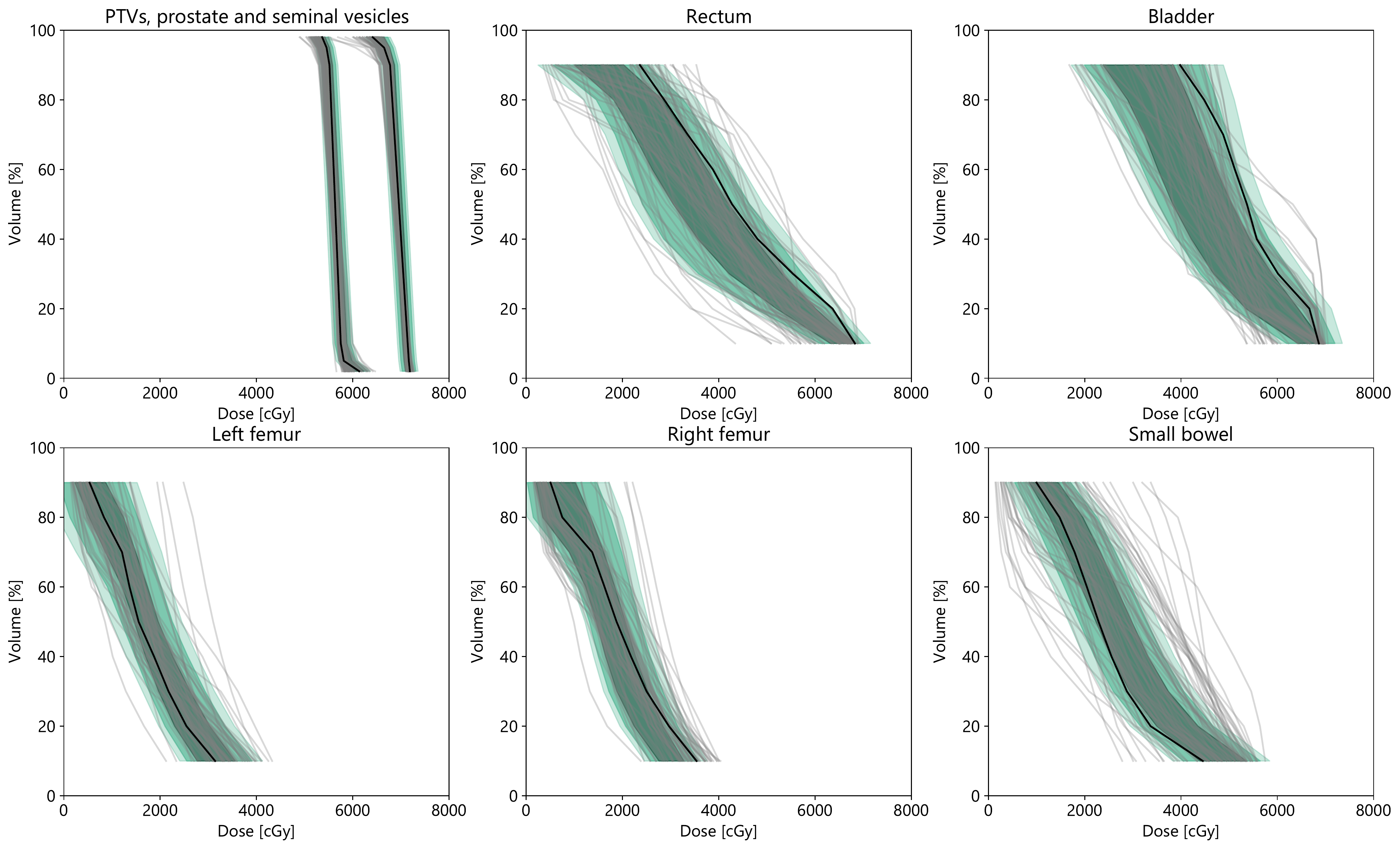}
\caption{Patient 1.}
\end{subfigure}%
\vspace{0.5cm}
\begin{subfigure}[b]{\textwidth}
\centering
\includegraphics[width=\textwidth]{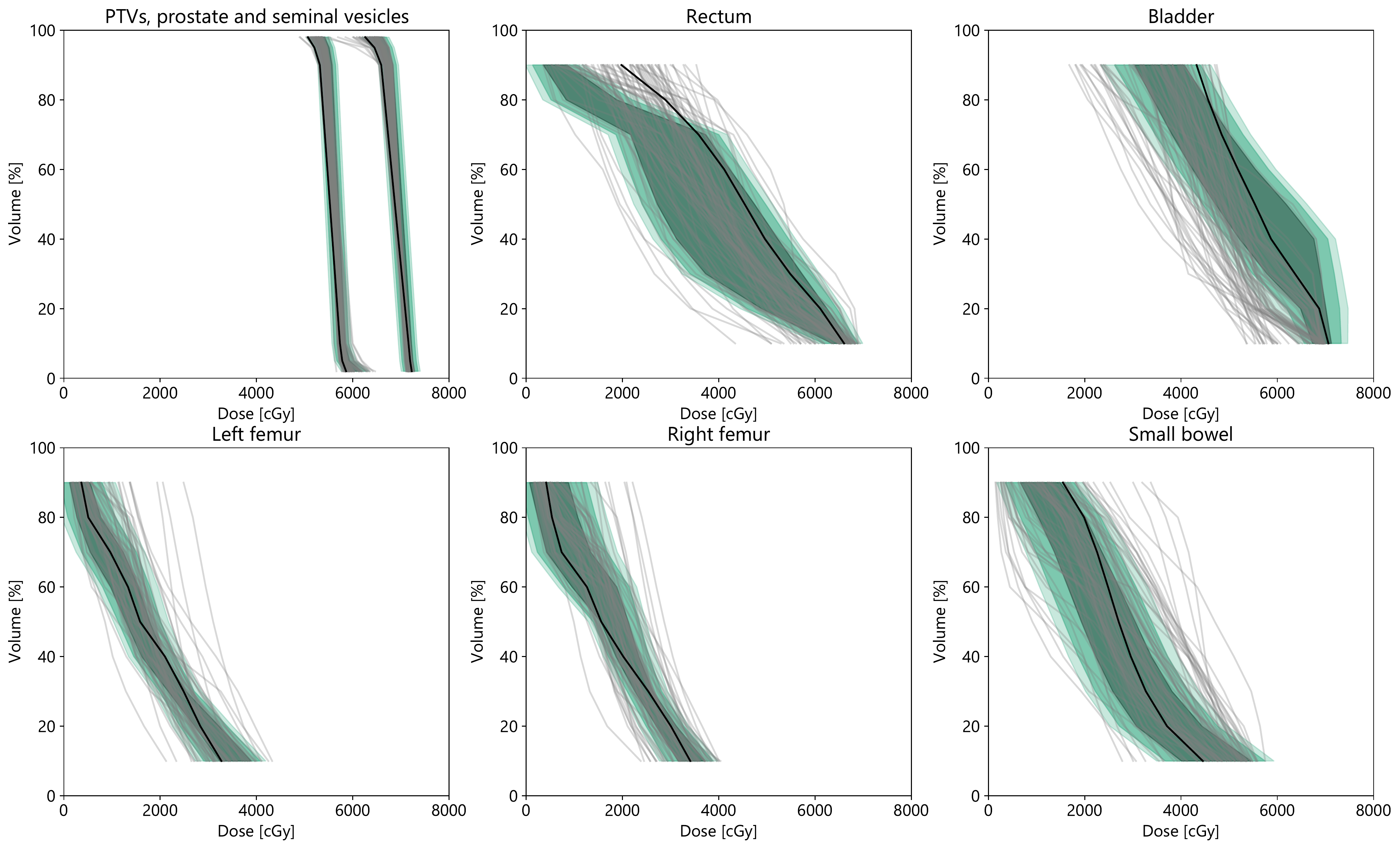}
\caption{Patient 2.}
\end{subfigure}
\caption{Illustration of dose statistic predictions for two test patients, where the shaded bands correspond to $99 \; \%$, $95 \; \%$ and $70 \; \%$ prediction intervals for each dose statistic. The training equivalents are shown in gray and the ground truth in black.}
\label{dvhband_1}
\end{figure}

Next, we assess the predictive performance of the seven ROI-specific dose statistic prediction models. Naturally, since the true predictive distributions are unknown, we cannot directly evaluate the accuracy of the estimated predictive distributions by computing e.g. Kullback--Leibler divergences to ground truths. However, we may use the training dataset as a reference sample and compare average errors between that and a sample drawn from the estimated predictive distribution. Table \ref{tableresults} shows the mean squared error (MSE) $\operatorname{\mathbb{E}} \Vert y - y^* \Vert^2_{\mathrm{std}}$ and the mean absolute error (MAE) $\operatorname{\mathbb{E}} \Vert y - y^* \Vert_{\mathrm{std}}$ for each model, averaged over the $10$ test patients---here, expectation is taken over $y \sim p(y \; | \; \phi(x^*), \{(\phi(x^n), y^n)\}_n)$ and $\Vert \cdot \Vert_{\mathrm{std}}$ denotes a standardized Euclidean norm using the sample standard deviations in the training dataset. Figure \ref{metric_boxplot} shows boxplots illustrating the spreads of the MSE and MAE errors over the test dataset beside corresponding benchmarks consisting of their empirical analogues over the training dataset. For the organs at risk, the dose statistic prediction outperforms the choice of a training data point at random, as would be expected---in particular, the interpretation of the figure is that the models mostly output predictions better than most other training data points, since the model boxes are centered around the lower half of their respective benchmark boxes. In this context, it is important to note that there is a theoretical limit on how low the boxplots may lie due to the inherent uncertainty of decisions made during treatment planning. For the targets, however, interestingly, the models perform slightly worse than the benchmark. There may be various reasons for this---for example, the predictive uncertainties may have been overestimated, or the target DVHs may not actually depend noticeably on the patient geometry. Finally, in Figure \ref{dvhband_1}, we show, for two patients in the test dataset, pointwise DVH confidence bands associated with the dose statistic prediction models for the seven ROIs. 

\begin{table}[h]
\caption{Comparison of the standardized MSE and MAE values, averaged over the test dataset, between the proposed models and their respective benchmarks, with the benchmark value defined as the average error over the training dataset. Note that the associated spreads of the validation metric values correspond to the boxplots in Figure \ref{metric_boxplot}.}
\centering
\begin{tabular}{lrrrr}
\toprule
\multirow{2}{*}[-5pt]{ROI} & \multicolumn{2}{c}{MSE} & \multicolumn{2}{c}{MAE}\\
\addlinespace[0pt]
\cmidrule(lr){2-3} \cmidrule(lr){4-5} \\
\addlinespace[-10pt]
{} & Model & Benchmark & Model & Benchmark \\
\midrule
PTV, prostate & $77.51$ & $21.49$ & $22.13$ & $12.62$ \\
PTV, seminal vesicles & $42.60$ & $25.78$ & $17.10$ & $13.22$ \\
Rectum & $13.11$ & $14.84$ & $8.54$ & $9.04$ \\
Bladder & $13.15$ & $18.51$ & $8.37$ & $9.97$ \\
Left femur & $11.67$ & $16.77$ & $7.69$ & $9.39$ \\
Right femur & $12.11$ & $16.06$ & $8.10$ & $9.09$ \\
Small bowel & $11.61$ & $15.12$ & $8.26$ & $9.21$ \\
\bottomrule
\end{tabular}
\label{tableresults}
\end{table}

\begin{figure}[h]
\centering
\begin{subfigure}[t]{\textwidth}
\centering
\includegraphics[width=\textwidth]{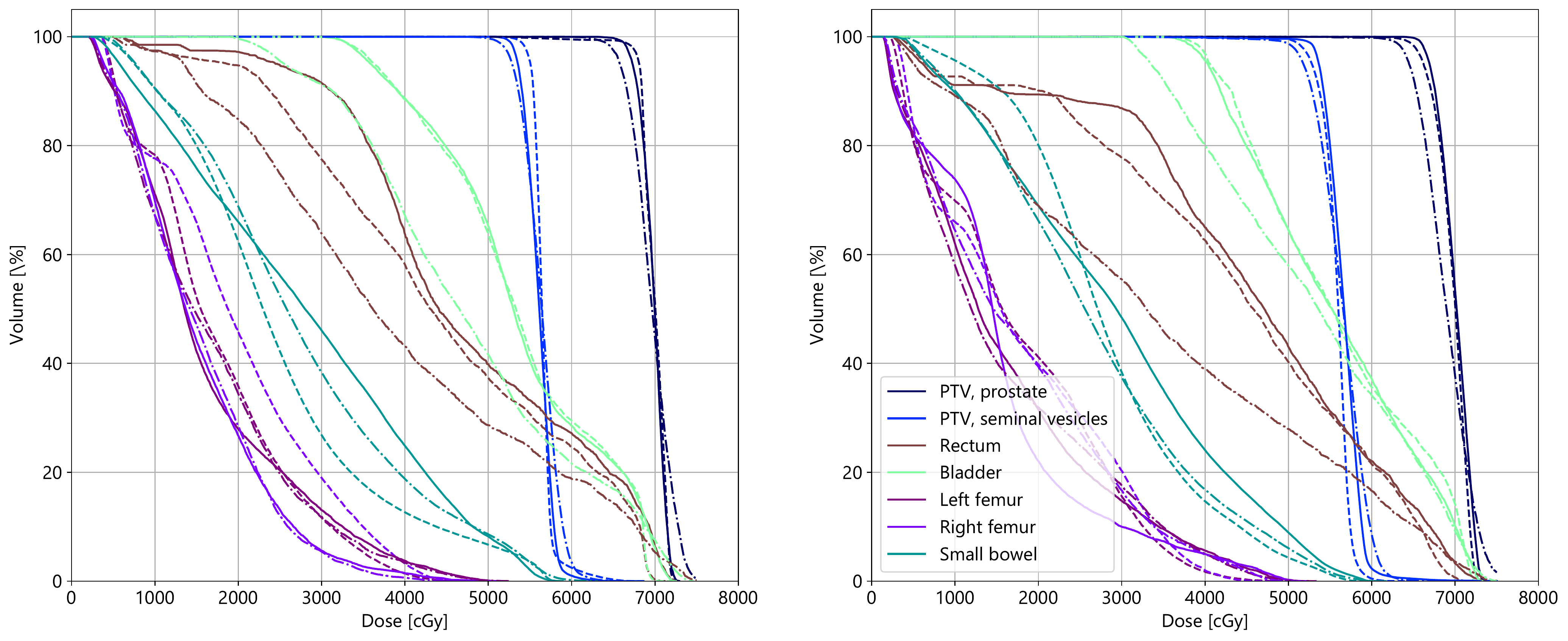}
\caption{}
\end{subfigure}%
\vspace{0.5cm}
\begin{subfigure}[b]{\textwidth}
\centering
\includegraphics[width=\textwidth]{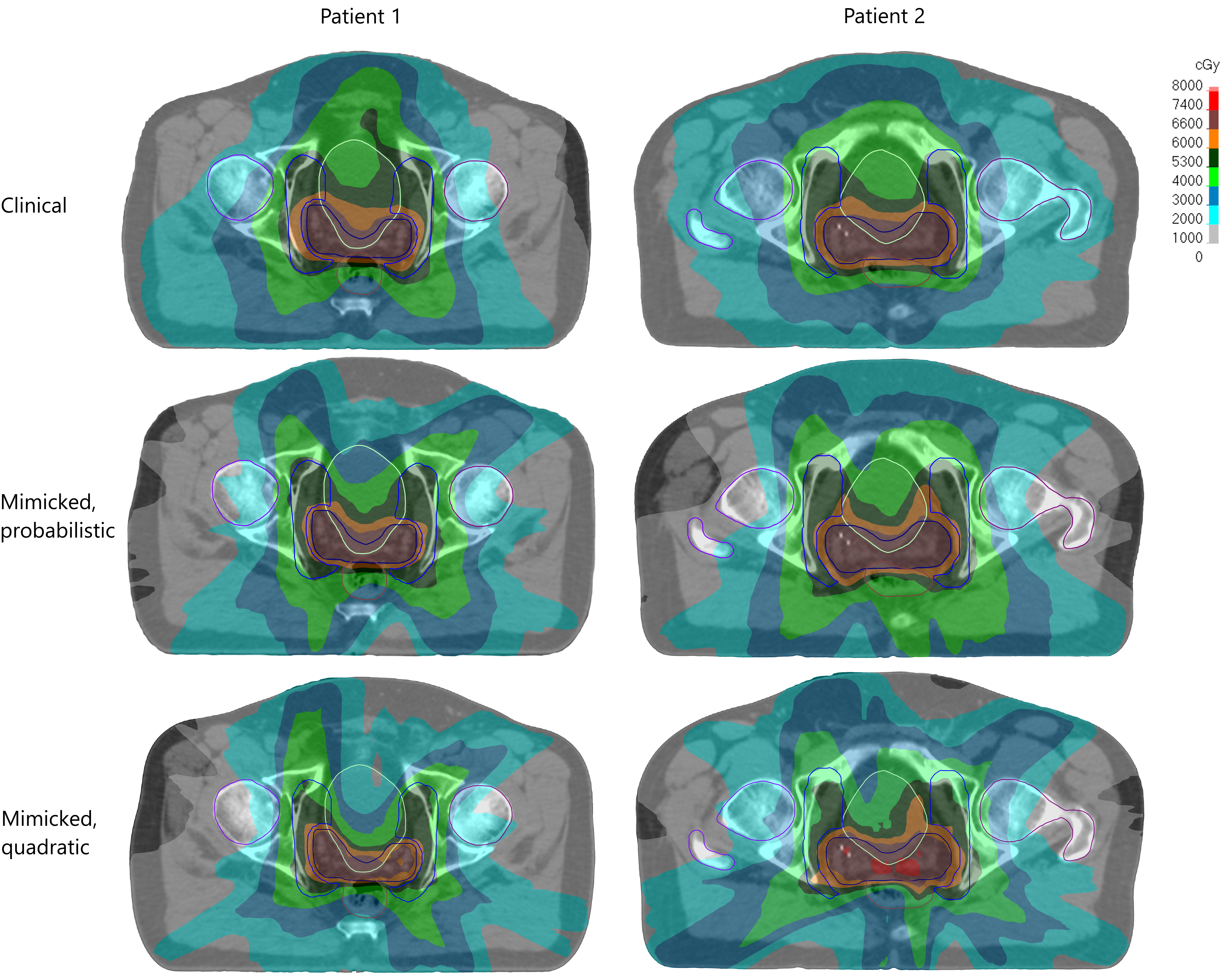}
\caption{}
\end{subfigure}
\caption{(a) DVH comparisons between the mimicked plans using the probabilistic (solid) and non-probabilistic (dash-dotted) formulations and their clinical counterparts (dashed), shown for the two test patients in Figure \ref{dvhband_1}. Note that the optimization problem is not set up to mimic the clinical plan but rather the predicted dose statistics. (b) Transversal cuts of the spatial dose distributions of the mimicked plans in comparison to their clinical counterparts, shown for the two test patients in Figure \ref{dvhband_1}. The colors of the ROI contours correspond to those in the legend in (a).}
\label{mimic_dvhs}
\end{figure}

Finally, to demonstrate that the dose mimicking method outlined in Section \ref{dosemimickingsec} can leverage the information contained in the predictive distributions from the dose statistic prediction, we create deliverable plans for two patients in the test dataset. This is then compared to the non-probabilistic formulation using quadratic penalties instead of cross-entropy objectives. The resulting DVHs and spatial dose distributions of the mimicked dose distributions are shown in Figure \ref{mimic_dvhs} in comparison to their ground truth counterparts. Interestingly, we observe that for both patient cases, the probabilistically mimicked DVHs follow the ground truth mostly well, possibly with slight deviations for the targets. On the other hand, the quadratic penalty mimicking was overly aggressive in reducing the dose to the rectum and the bladder at the cost of sacrificing target coverage and homogeneity. The reason for this is that the quadratic penalties, containing no uncertainty insight, are overconfident that the predictive mean DVHs could be achieved---the predictions in the rectum and the bladder, while being colder than the clinical plan for both patient cases, are associated with a much higher uncertainty than for the targets, which the probabilistic dose mimicking formulation was able to take into account. The specification of preferences captured in the predictive distributions are thus being realized differently depending on the actual patient geometry. Hence, this showcases the advantage of articulating one's preferences by a probability distribution instead of a single reference value. On the other hand, we see that the spatial dose distributions for both mimicking formulations follow the ground truths mostly well but may differ significantly in parts not covered by any ROI, with the quadratic penalty formulation deviating slightly more, which is expected since the optimization problem does not penalize such deviations. While we may conclude that the comparison demonstrates the merits of a probabilistic approach to dose mimicking, it is emphasized that in a clinical setting, further post-processing is in general required to obtain plans of quality sufficient to be approved and delivered.

\section{Discussion}

In the context of predicting dose-related quantities as part of an automated treatment planning pipeline, the importance of precise quantification of predictive uncertainties has been left largely unaddressed in literature. Motivated by this, in this work, we have presented a general framework in which the multivariate predictive distribution of any collection of dose-related quantities may be estimated and leveraged in a dose mimicking problem. We have shown that a variational autoencoder can be employed to extract from contoured CT images lower-dimensional features of substantial relevance for predicting dose statistics, and that a similarity-based mixture-of-experts model can be used to quantify predictive uncertainties of dose statistic values, including inter-statistic dependencies. Moreover, we have shown that by setting up a dose mimicking problem in which contributions from dose statistics are based on the marginal cumulative distribution functions in a cross-entropy--like form, with gradients of dose statistics directly available by virtue of recent developments, we may create deliverable plans agreeing well with their respective historical clinical plans with minimal need for weight tuning. In particular, the comparison between the proposed dose mimicking formulation and a non-probabilistic analogue has showcased the merits of including uncertainty information during optimization. 

The nonparametric mixture-of-experts method used for dose statistic prediction is similar to previous work in terms of training requirements, portability and generalizability. Compared to the methods in early studies on knowledge-based planning \citep{zhu, wu, yuan}, instead of predicting principal component coefficients, our method outputs complete probability distributions capable of modeling inter-statistic dependencies, skewness and eventual multimodality. As argued by \citet{babier_importance}, it is crucial to maintain a holistic perspective when developing and assessing constituent methods in a prediction--mimicking pipeline---there is, for instance, no direct causality between the performance of a dose prediction model and the quality of the produced plan. In this sense, the proposed division of the pipeline into feature extraction, dose statistic prediction and dose mimicking has the advantages of being flexible, with each part being substitutable with any other algorithm, and general, with minimal loss of information between the steps. Moreover, the former two parts may naturally be combined into a semi-supervised learning algorithm, reducing the need for training data in the form of clinically delivered treatments. A disadvantage, however, is that the dose-related quantities to be predicted must be meaningful when evaluated on all training patients, which is not true for e.g. individual voxel doses of the current patient. Thus, a spatial dose prediction algorithm may complement the current dose mimicking problem setup by helping to shape the dose distribution outside the defined ROIs. If done also probabilistically, such as in \citet{nguyen_bagging}, the spatial dose prediction could be basis for a contribution to the dose mimicking objective in (\ref{optformulation}) penalizing spatial deviation, similarly to in \citet{nilsson}, as well as serve as a reference for our proposed dose statistic prediction method.

While the present work primarily focuses on conceptually demonstrating the merits of the proposed method, there are several interesting directions for future research. For example, the features extracted by the variational autoencoder may benefit from being augmented with handcrafted counterparts such as in \citet{cui}, and the estimated predictive distribution over dose statistics may be used for quality assurance purposes. Furthermore, with an estimated joint predictive distribution, one may use the inter-statistic dependencies to post-process the distribution to be more aggressive toward one or several dose statistics. This may, in turn, be used to form differently focused tradeoff objectives in a multicriteria optimization problem, leading to a semiautomated treatment planning workflow in which the user may optionally articulate their own preferences by navigating between Pareto optimal plans. In conclusion, we note that a rigorous probabilistic perspective of the automated treatment planning problem enables ample new opportunities to be further explored in future work. 

\section{Conclusions}

In this work, we have presented a general framework for automated radiation therapy treatment planning based on a three-step pipeline comprising feature extraction, dose statistic prediction and dose mimicking. By using latent representations produced by a variational autoencoder and a nonparametric Bayesian dose statistic prediction method, we obtain reasonable predictive distributions as multivariate Gaussian mixtures. Furthermore, using objective functions specially designed to leverage such predictive information, we perform dose mimicking to create deliverable plans resembling their original counterparts well. Thus, we have demonstrated the advantages of a holistic approach based on probabilistic modeling, with applications to many related tasks.

\printbibliography

@article{ge,
	author   = {Yaorong Ge and Q Jackie Wu},
	title    = {Knowledge-based planning for intensity-modulated radiation therapy: a review of data-driven approaches},
	journal  = {Med. Phys.},
	volume   = {46},
	number   = {6},
	pages    = {2760--2775},
	year     = {2019},
}

@article{boutilier,
	author   = {Justin J Boutilier and Taewoo Lee and Tim Craig and Michael B Sharpe and Timothy C Y Chan},
	title    = {Models for predicting objective function weights in prostate cancer IMRT},
	journal  = {Med. Phys.},
	volume   = {42},
	number   = {4},
	pages    = {1586--1595},
	year     = {2015},
}

@article{shen,
	author   = {Chenyang Shen and Dan Nguyen and Liyuan Chen and Yesenia Gonzalez and Rafe McBeth and Nan Qin and Steve B Jiang and Xun Jia},
	title    = {Operating a treatment planning system using a deep‐reinforcement learning‐based virtual treatment planner for prostate cancer intensity‐modulated radiation therapy treatment planning},
	journal  = {Med. Phys.},
	volume   = {64},
	number   = {4},
	pages    = {115013},
	year     = {2020},
}

@article{park,
	author   = {Cheolsoo Park and Clive Cheong Took and Joon-Kyung Seong},
	title    = {Machine learning in biomedical engineering},
	journal  = {Biomed. Eng. Lett.},
	volume   = {8},
	pages    = {1--3},
	year     = {2018},
}

@article{babier_importance,
	author   = {Aaron Babier and Rafid Mahmood and Andrea L McNiven and Adam Diamant and Timothy C Y Chan},
	title    = {The importance of evaluating the complete automated knowledge-based planning pipeline},
	journal  = {Phys. Med.},
	volume   = {72},
	pages    = {73--79},
	year     = {2020},
}

@article{siddique,
	author   = {Siddique, Sarkar and Chow, James C L},
	title    = {Artificial intelligence in radiotherapy},
	journal  = {Rep. Pract. Oncol. Radiother.},
	volume   = {25},
	number   = {4},
	pages    = {656--666},
	year     = {2020},
}

@article{ng,
	author   = {Ng, Frederick and Jiang, Runqing and Chow, James C L},
	title    = {Predicting radiation treatment planning evaluation parameter using artificial intelligence and machine learning},
	journal  = {IOP SciNotes},
	volume   = {1},
	number   = {1},
	pages    = {014003},
	year     = {2020},
}

@article{wang,
	author   = {Chunhao Wang and Xiaofeng Zhu and Julian C Hong and Dandan Zheng},
	title    = {Artificial intelligence in radiotherapy
treatment planning: present and future},
	journal  = {Technol. Cancer Res. Treat.},
	volume   = {18},
	pages    = {1--11},
	year     = {2019},
}

@article{hussein,
	author   = {Mohammad Hussein and Ben J M Heijmen and Dirk Verellen and Andrew Nisbet},
	title    = {Automation in intensity modulated radiotherapy
treatment planning---a review of recent innovations},
	journal  = {Br. J. Radiol.},
	volume   = {91},
	pages    = {20180270},
	year     = {2018},
}

@article{nguyen_unet,
    author = {Dan Nguyen and Troy Long and Xun Jia and Weiguo Lu and Xuejun Gu and Zohaib Iqbal and Shucui Jiang},
    year = {2019},
    volume={9},
    number={1079},
    journal={Sci. Rep.},
    title = {A feasibility study for predicting optimal radiation therapy dose distributions of prostate cancer patients from patient anatomy using deep learning}
}

@article{campbell,
    title={Neural network dose models for knowledge-based planning in pancreatic SBRT},
    author={Campbell, Warren G and Miften, Moyed and Olsen, Lindsey and Stumpf, Priscilla and Schefter, Tracey and Goodman, Karyn A and Jones, Bernard L},
    journal={Med. Phys.},
    volume={44},
    number={12},
    pages={6148--6158},
    year={2017}
}

@article{kearney,
    title={DoseNet: a volumetric dose prediction algorithm using 3D fully-convolutional neural networks},
    author={Kearney, Vasant and Chan, Jason W and Haaf, Samuel and Descovich, Martina and Solberg, Timothy D},
    journal={Phys. Med. Biol.},
    volume={63},
    number={23},
    pages={235022},
    year={2018}
}

@article{shiraishi,
  title={Knowledge-based prediction of three-dimensional dose distributions for external beam radiotherapy},
  author={Shiraishi, Satomi and Moore, Kevin L},
  journal={Med. Phys.},
  volume={43},
  number={1},
  pages={378--387},
  year={2016}
}

@article{ma_isodose,
	author   = {Ming Ma and Mark K Buyyounouski and Varun Vasudevan and Lei Xing and Yong Yang},
	title    = {Dose distribution prediction in isodose feature-preserving voxelization domain using deep convolutional neural network},
	journal  = {Med. Phys.},
	volume   = {46},
	number   = {7},
	pages    = {2978--2987},
	year     = {2019},
}

@article{babier_gan,
  title={Knowledge-based automated planning with three-dimensional generative adversarial networks},
  author={Babier, Aaron and Mahmood, Rafid and McNiven, Andrea L and Diamant, Adam and Chan, Timothy C Y},
  journal={Med. Phys.},
  volume={47},
  number={2},
  pages={297--306},
  year={2020},
  publisher={Wiley Online Library}
}

@article{murakami,
    title={Fully automated dose prediction using generative adversarial networks in prostate cancer patients},
    author={Murakami, Yu and Magome, Taiki and Matsumoto, Kazuki and Sato, Tomoharu and Yoshioka, Yasuo and Oguchi, Masahiko},
    journal={PLoS One},
    volume={15},
    number={5},
    pages={e0232697},
    year={2020}
}

@article{mcintosh,
	author   = {Chris McIntosh and Mattea Welch and Andrea McNiven and David A Jaffray and Thomas G Purdie},
	title    = {Fully automated treatment planning for head and neck radiotherapy using a voxel-based dose prediction and dose mimicking method},
	journal  = {Phys. Med. Biol.},
	volume   = {62},
	number   = {15},
	pages    = {5926--5944},
	year     = {2017},
}

@article{appenzoller,
	author   = {Lindsey M Appenzoller and Jeff M Michalski and Wade L Thorstad and Sasa Mutic and Kevin L Moore},
	title    = {Predicting dose--volume histograms for organs-at-risk in IMRT planning},
	journal  = {Med. Phys.},
	volume   = {39},
	number   = {12},
	pages    = {7446--7461},
	year     = {2012},
}

@article{liu,
	author   = {Zhiqiang Liu and Xinyuan Chen and Kuo Men and Junlin Yi and Jianrong Dai},
	title    = {A deep learning model to predict dose--volume histograms of organs at risk in radiotherapy treatment plans},
	journal  = {Med. Phys.},
	volume   = {47},
	number   = {11},
	pages    = {5467--5481},
	year     = {2020},
}

@article{nguyen_pareto,
	author   = {Dan Nguyen and Rafe McBeth and Azar Sadeghnejad Barkousaraie and Gyanendra Bohara and Chenyang Shen and Xun Jia and Steve Jiang},
	title    = {Incorporating human and learned domain knowledge into training deep neural networks: a differentiable dose--volume histogram and adversarial inspired framework for generating Pareto optimal dose distributions in radiation therapy},
	journal  = {Med. Phys.},
	volume   = {47},
	number   = {3},
	pages    = {837--849},
	year     = {2020},
}

@article{ma_features,
	author   = {Ming Ma and Nataliya Kovalchuk and Mark K Buyyounouski and Lei Xing and Yong Yang},
	title    = {Dosimetric features--driven machine learning model for DVH prediction in VMAT treatment planning},
	journal  = {Med. Phys.},
	volume   = {46},
	number   = {2},
	pages    = {857--867},
	year     = {2019},
}

@article{wall,
	author   = {Phillip D H Wall and Robert L Carver and Jonas D Fontenot},
	title    = {An improved distance-to-dose correlation for predicting bladder and rectum dose--volumes in knowledge-based VMAT planning for prostate cancer},
	journal  = {Phys. Med. Biol.},
	volume   = {63},
	pages    = {015035},
	year     = {2018},
}

@article{jiao,
	author   = {Sheng-Xiu Jiao and Li-Xin Chen and Jin-Han Zhu and Ming-Li Wang and Xiao-Wei Liu},
	title    = {Prediction of dose--volume histograms in nasopharyngeal cancer IMRT using geometric and dosimetric information},
	journal  = {Phys. Med. Biol.},
	volume   = {64},
	pages    = {23NT04},
	year     = {2019},
}

@article{skarpmanmunter,
	author   = {Skarpman Munter, Johanna and Sjölund, Jens},
	title    = {Dose--volume histogram prediction using density estimation},
	journal  = {Phys. Med. Biol.},
	volume   = {60},
	number   = {17},
	pages    = {6923--6936},
	year     = {2015},
}

@article{nilsson,
	author   = {Nilsson, Viktor and Gruselius, Hanna and Zhang, Tianfang and De Kerf, Geert and Claessens, Michaël},
	title    = {Probabilistic dose prediction using mixture density networks for automated radiation therapy treatment planning},
	journal  = {Phys. Med. Biol.},
	volume   = {66},
	number   = {5},
	pages    = {055003},
	year     = {2021},
}

@article{fu,
	author   = {Yabo Fu and Yang Lei and Tonghe Wang and Walter J Curran and Tian Liu and Xiaofeng Yang},
	title    = {Deep learning in medical image registration: a review},
	journal  = {Phys. Med. Biol.},
	volume   = {65},
	number   = {20},
	pages    = {20TR01},
	year     = {2020},
}

@article{cui,
	author   = {Cui, Sunan and Luo, Yi and Tseng, Huan-Hsin Tseng and Ten Haken, Randall K and El Naqa, Issam},
	title    = {Combining handcrafted features with latent variables in machine learning for prediction of radiation-induced lung damage},
	journal  = {Med. Phys.},
	volume   = {46},
	number   = {5},
	pages    = {2497--2511},
	year     = {2019},
}

@article{bengio,
	author   = {Yoshua Bengio and Aaron Courville and Pascal Vincent},
	title    = {Representation learning: a review and new perspectives},
	journal  = {IEEE Trans. Pattern Anal. Mach. Intell.},
	volume   = {35},
	number   = {8},
	pages    = {1798--1828},
	year     = {2013},
}

@book{goodfellow,
    title={Deep Learning},
    author={Ian Goodfellow and Yoshua Bengio and Aaron Courville},
    publisher={MIT Press},
    year={2016}
}

@article{blei,
	author   = {David M Blei and Alp Kucukelbir and Jon D McAuliffe},
	title    = {Variational inference: a review for statisticians},
	journal  = {J. Am. Stat. Assoc.},
	volume   = {112},
	number   = {518},
	pages    = {859--877},
	year     = {2017},
}

@inproceedings{kingma,
	author   = {Diederik P Kingma and Max Welling},
	title    = {Auto-encoding variational Bayes},
	booktitle  = {International Conference on Learning Representations},
	year     = {2014},
}

@misc{zhang_sbmoe,
    title={A similarity-based Bayesian mixture-of-experts model},
    author={Tianfang Zhang and Rasmus Bokrantz and Jimmy Olsson},
    year={2020},
    eprint={2012.02130},
    archivePrefix={arXiv},
    primaryClass={stat.ML}
}

@article{unkelbach,
	author   = {J Unkelbach and T Bortfeld and D Craft and M Alber and M Bangert and R Bokrantz and D Chen and R Li and L Xing and C Men and S Nill and D Papp and H E Romeijn and E Salari},
	title    = {Optimization approaches to volumetric modulated arc therapy planning},
	journal  = {Med. Phys.},
	volume   = {42},
	number   = {3},
	pages    = {1367--1377},
	year     = {2015},
}

@article{gay,
    title = "Pelvic normal tissue contouring guidelines for radiation therapy: a radiation therapy oncology group consensus panel atlas",
    journal = "Int. J. Radiat. Oncol. Biol. Phys.",
    volume = "83",
    number = "3",
    pages = "e353--362",
    year = "2012",
    author = "Hiram A. Gay and H. Joseph Barthold and Elizabeth O’Meara and Walter R. Bosch and Issam {El Naqa} and Rawan Al-Lozi and Seth A. Rosenthal and Colleen Lawton and W. Robert Lee and Howard Sandler and Anthony Zietman and Robert Myerson and Laura A. Dawson and Christopher Willett and Lisa A. Kachnic and Anuja Jhingran and Lorraine Portelance and Janice Ryu and William Small and David Gaffney and Akila N. Viswanathan and Jeff M. Michalski",
}

@article{salembier,
    title = "ESTRO ACROP consensus guideline on CT- and MRI-based target volume delineation for primary radiation therapy of localized prostate cancer",
    journal = "Radiother. Oncol.",
    volume = "127",
    number = "1",
    pages = "49--61",
    year = "2018",
    author = "Carl Salembier and Geert Villeirs and Berardino {De Bari} and Peter Hoskin and Bradley R. Pieters and Marco {Van Vulpen} and Vincent Khoo and Ann Henry and Alberto Bossi and Gert {De Meerleer} and Valérie Fonteyne",
}

@misc{kingma_adam,
    author = "Diederik P Kingma and Jimmy Ba",
    title = "Adam: a method for stochastic optimization",
    year = 2014, 
    eprint={1412.6980},
    archivePrefix={arXiv},
    primaryClass={cs.LG}
}

@article{zhang_direct,
	author   = {Tianfang Zhang and Rasmus Bokrantz and Jimmy Olsson},
	title    = {Direct optimization of dose--volume histogram metrics in radiation therapy treatment planning},
	journal  = {Biomed. Phys. Eng. Express},
	volume   = {6},
	number   = {6},
	pages    = {065018},
	year     = {2020},
}

@article{romeijn,
	author   = {H Edwin Romeijn and Ravindra K Ahuja and James F Dempsey and Arvind Kumar},
	title    = {A new linear programming approach to radiation therapy treatment planning problems},
	journal  = {Oper. Res.},
	volume   = {54},
	pages    = {201--216},
	year     = {2006},
}

@mastersthesis{hanna,
    author    = {Hanna Gruselius},
    title     = {Generative models and feature extraction on patient images and structure data in radiation therapy},
    school    = {KTH Royal Institute of Technology},
    year      = {2018},
}

@article{nelms,
	author   = {Benjamin E Nelms and Greg Robinson and Jay Markham and Kyle Velasco and Steve Boyd and Sharath Narayan and James Wheeler and Mark L Sobczak},
	title    = {Variation in external beam treatment plan quality: an inter-institutional study of planners and planning systems},
	journal  = {Pract. Radiat. Oncol.},
	volume   = {2},
	number   = {4},
	pages    = {296--305},
	year     = {2012},
}

@article{nguyen_bagging,
	author   = {Nguyen, Dan and Barkousaraie, Azar Sadeghnejad and Bohara, Gyanendra and Balagopal, Anjali and McBeth, Rafe and Lin, Mu-Han and Jiang, Steve},
	title    = {A comparison of Monte Carlo dropout and bootstrap aggregation on the performance and uncertainty estimation in radiation therapy dose prediction with deep learning neural networks},
	journal  = {Phys. Med. Biol.},
	volume   = {66},
	number   = {5},
	pages    = {054002},
	year     = {2021},
}

@article{zhu,
	author   = {Zhu, Xiaofeng and Ge, Yaorong and Li, Taoran and Thongphiew, Danthai and Yin, Fang-Fang and Wu, Q Jackie},
	title    = {A planning quality evaluation tool for prostate adaptive IMRT based on machine learning},
	journal  = {Med. Phys.},
	volume   = {38},
	number   = {2},
	pages    = {719--726},
	year     = {2011},
}

@article{wu,
	author   = {Wu, Binbin and Ricchetti, Francesco and Sanguineti, Giuseppe and Kazhdan, Michael and Simari, Patricio and Jacques, Robert and Taylor, Russell and McNutt, Todd},
	title    = {Data-driven approach to generating achievable dose--volume histogram objectives in intensity-modulated radiotherapy planning},
	journal  = {Int. J. Radiat. Oncol. Biol. Phys.},
	volume   = {79},
	number   = {4},
	pages    = {1241--1247},
	year     = {2011},
}

@article{yuan,
	author   = {Yuan, Lulin and Yaorong, Ge and Lee, W Robert and Yin, Fang Fang and Kirkpatrick, John P and Wu, Q Jackie},
	title    = {Quantitative analysis of the factors which affect the interpatient organ-at-risk dose sparing variation in IMRT plans},
	journal  = {Med. Phys.},
	volume   = {39},
	number   = {11},
	pages    = {6868--6878},
	year     = {2012},
}

@article{fogliata,
	author   = {Fogliata, A and Cozzi, L and Reggiori, G and Stravato, A and Lobefalo, F and Franzese, C and Franceschini, D and Tomatis, S and Scorsetti, M},
	title    = {RapidPlan knowledge based planning: iterative learning process and model ability to steer planning strategies},
	journal  = {Radiat. Oncol.},
	volume   = {14},
	pages    = {187},
	year     = {2019},
}

\end{document}